\documentclass[a4paper,11pt]{article}
\usepackage{authblk}
\usepackage[utf8]{inputenc}
\usepackage[british]{babel}
\usepackage[mono=false]{libertine}
\usepackage[T1]{fontenc}
\usepackage[top=2.5cm, bottom=2.5cm, left=2cm, right=2cm]{geometry}
\usepackage[font=small]{caption}
\usepackage{subcaption}
\usepackage{cite}
\usepackage{enumitem}
\usepackage{titlesec}
\usepackage[usenames, dvipsnames]{color}
\usepackage[titletoc]{appendix}
\usepackage[pdftex]{graphicx}
\usepackage{array}
\usepackage{url}
\usepackage{mathtools}
\usepackage{amssymb,amsfonts,amsmath,amsthm}
\usepackage{dsfont}
\usepackage{bm}
\usepackage[perpage,hang,flushmargin]{footmisc} 
\usepackage{enumitem}
\usepackage{booktabs}       
\usepackage{nicefrac}       
\usepackage{microtype}      
\usepackage{tikz}
\usepackage[ruled,vlined]{algorithm2e}
\usepackage{wrapfig}
\usepackage{xr,xr-hyper}
\usepackage{cases}
\usepackage{setspace}
\usepackage[pdfpagemode=UseNone,bookmarksopen=false,colorlinks=true,urlcolor=blue,citecolor=magenta,citebordercolor=blue,linkcolor=blue]{hyperref}
\usepackage{cleveref}
\makeatletter
\newcommand{\setappendix}{Appendix~\thesection:~~}
\newcommand{\setsection}{\thesection~~}
\titleformat{\section}{\bfseries\LARGE}{%
	\ifnum\pdfstrcmp{\@currenvir}{appendices}=0
	\setappendix
	\else
	\setsection
\fi}{0em}{}
\makeatother
\DeclareUnicodeCharacter{00A0}{~}

\newcommand{\cJ}{{\mathcal {J}}}

\theoremstyle{definition}

\makeatletter
\renewcommand\paragraph{\@startsection{paragraph}{4}{\z@}%
                                      {\parskip}
                                      {-1em}%
                                      {\normalfont\normalsize\bfseries}}
\makeatother

\newcommand{\bbR}{\mathbb{R}}

\newcommand{\cP}{\mathcal{P}}
\newcommand{\cD}{\mathcal{D}}
\allowdisplaybreaks

\begin{document}

\setcounter{tocdepth}{2}
\setcounter{secnumdepth}{3}

\title{
Spin glass theory and its new challenge:
  structured disorder}

\date{\today}
\author[$\star$]{ Marc M\'ezard
}
\affil[$\star$]{ Bocconi University, Milano, Italy
  }
\maketitle

\begin{abstract}
This paper first describes, from a high level viewpoint, the main
challenges that had to be solved in order to develop a theory of spin
glasses in the last fifty years. It then explains how important
inference problems, notably those occurring in machine learning, can
be formulated as problems in statistical physics of disordered
systems. However, the main questions that we face in the analysis of
deep networks require to develop a new chapter of spin glass theory,
which will address the  challenge of structured data. 
\end{abstract}

\begin{spacing}{1.0}
\tableofcontents
\end{spacing}

\renewcommand{\labelitemi}{$\bullet$}

\section{Spin glasses}

Statistical physics is more or less one and a half century old. Its creation was based on renouncing to follow the
trajectories of single particles and moving rather to a coarser,
statistical description of systems with many interacting
particles. This radical move allowed to handle the specific effects
that emerge when the number of particles becomes large, as summarized
in Phil Anderson's famous paper ``More is
different''\cite{andersonMore}. One of its great achievements is
the understanding and analysis of phase transitions, and the
discovery of universality classes at second order phase transitions,
where the divergence of the correlation length wipes out many of the
microscopic details of the particles.

About fifty years ago, statistical physics developed a new research direction,
the one of strongly disordered systems. An important building piece of
its construction is the theory of spin glasses, magnetic systems with
disordered interactions. In this section we shall mention some of the 
formidable challenges that had to be solved in order to develop a
theory of spin glasses, keeping to the case of classical systems
(parallel developments in the field of quantum statistical
physics  deserve a separate presentation).

As is well known, magnetism has played an important role in the development of
statistical physics. The solution by Onsager of a ``simple'' model of ferromagnet, the Ising
model in two dimensions, was crucial in establishing the concept of
spontaneous symmetry breaking. And the understanding of Ising models in $d$-dimensions,
including non-integer values of $d$, also played a major rule in the
development of the renormalization group.

\subsection{A few well known landmarks from the ferromagnetic Ising model}
It is useful to set the stage and prepare the discussion of spin
glasses, starting with a very short sketch of the well known case of
ferromagnetism, which can be found in any standard book of statistical
physics. In the Ising model, two-states spins $s_i=\pm 1$ located on the
$N=L^d$ vertices of a hypercubic
$d$-dimensional lattice interact through pair interactions, with an
interaction energy which is a sum over pairs of adjacent spins
\begin{align}
  E(s)=-J\sum_{(ij)} s_i s_j
  \label{eq:Ising}
\end{align}
where $J>0$ is the ferromagnetic coupling constant.

The order parameter is the magnetization density
\begin{align}
  M=\frac{1}{N}\sum_i \langle s_i \rangle
  \label{eq:Mdef}
\end{align}
Where $\langle s_i \rangle$ is the expectation of the spin $s_i$ with
respect to the Boltzmann measure $P(s)=(1/Z) e^{-\beta
  E(s)}$. This measure is even by the simultaneous flipping of all the
spins $s_i\to -s_i$, therefore for a fixed $N$ one has $M=0$ at any
inverse temperature $\beta$. On the
other hand, if one adds a small symmetry breaking term to the energy,
$E(s)\to E(s) -B\sum_i s_i$, then
\begin{align}
  \lim_{B\to 0^{\pm}} \lim_{N\to \infty} M= \pm M^*
  \label{ferro-mag}
\end{align}
with $M^*(\beta)>0$ in the low temperature phase $\beta>\beta_c$,
where the inverse critical temperature $\beta_c$ is finite for $d\geq
2$. This is the ferromagnetic phase transition, associated with the phenomenon of spontaneous symmetry breaking.

In order to get a first qualitative understanding of this phase
transition, one can use the mean field approximation. Starting from
the exact relation
\begin{align}
\langle s_i \rangle=\langle \tanh \beta\left(B+J\sum_{j\in D_i} s_j\right) \rangle
\end{align}
where $D_i$ is the set of neighbours of spin $i$, one neglects fluctuations, substituting the expectation value of the
$\tanh$ by the  $\tanh$ of the expectation value (this is the mean field). Seeking a
homogeneous solution $\langle s_i\rangle =M$ (which is correct far
from the boundaries, or using periodic boundary conditions), one finds
\begin{align}
  M= \tanh\beta \left(B+z J M \right)
  \label{meanfield}
\end{align}
where $z=|D_i|$ is the number of neighbors of each spin. This equation
predicts a ferromagnetic phase transition when $B\to 0$, with an inverse critical
temperature $\beta_c=1/(z J)$.

The mean field approximation is better in larger dimensions, it
actually becomes exact when $d\to \infty$, while it wrongly predicts
the existence of a phase transition when $d=1$.
A popular model where
the mean field approximation becomes exact is the Curie Weiss model,
where all the pairs of spins interact with a rescaled coupling
$J=\tilde J/N$. In thie model  the mean field equation $ M= \tanh\beta
\left(B+\tilde J M \right)$ is exact and the phase transition takes
place at $\beta_c=1/\tilde J$.

\subsection{Spin glasses}
A simple model for spin glasses is the Edwards Anderson model\cite{EA}. This
has the same ingredients as the Ising model, except that the coupling
constant between two spins $i,j$ depends on the pair. The energy
becomes
\begin{align}
  E(s)=-\sum_{(ij)} J_{ij} s_i s_j
  \label{eq:EA}
\end{align}
Depending on the pair $(ij)$, the coupling constant can be
ferromagnetic ($J_{ij}>0$, favoring the alignment of spins at low
temperatures), or antiferromagnetic  ($J_{ij}>0$, favoring spins
pointing in opposite directions at low
temperatures). 

With respect to the ferromagnetic case this
modification is crucial, and poses a number of remarkable challenges
that had to be solved in order to elaborate a theory of spin
glasses. This elaboration is a remarkable achievement which culminated
in the solution by Parisi of the mean-field Sherrington-Kirkpatrick
model\cite{SK} (Parisi's Nobel lecture \cite{ParisiNobel} gives a nice
summary, and the recent book \cite{FarBeyond} gives an idea of the
applications that it has had in several branches of science).

In this paper, I shall not enter any detail of spin glass theory, but
adopt a high-level point of view, trying to point out the  fourmost important challenges.

\subsection{First Challenge: Ensembles of samples}
The first challenge that  can be identified is the characterization of a spin
glass sample. In order to define the energy, and therefore the Boltzmann
probability, one needs to know all the coupling constants
$\cJ=\{J_{ij}\}_{1\leq i<j\leq N}$. If
the interactions are short range, this is a number of parameters which
grows proportionnally to the size of the system $N$. This raises two
problems. On the one hand, for macroscopic systems  it is
impossible to even write the energy function: the description of a
given sample requires to know  a number of  parameters of the order
of the Avogadro number.
On the other hand, for each new sample characterized by these
couplings $\cJ$, there is a new Boltzmann probability
\begin{align}
  P_{\cJ}(s)=\frac{1}{Z_\cJ}e^{\beta \sum_{(ij)} J_{ij} s_i s_j}
  \end{align}

In a step that mimics the one which was taken when statistical physics
was first introduced, this double problem was solved by introducing a
second level of probability, namely a probability distribution in the
space of samples. The couplings $\cJ$ are supposed to be generated
from a probability distribution $\cP(\cJ)$. A given realization of
$\cJ$ is a sample. For instance in the Edwards-Anderson model
\cite{EA} one
assumes that for each pair $(ij)$ of neighboring spins we draw
$J_{ij}$ independently at random, from a distribution with probability
density $\rho$. In the SK model each of the $N(N-1)/2$ couplings
$J_{ij}$ is drawn at random from a normal distribution with mean $0$
and variance $1/N$.

We have now two levels of probability. First one draws a sample $\cJ$
generated from the probability $\cP(\cJ)$. Then, one studies the
Boltzmann law $P_\cJ(s)$ for this sample. The averages of spin
configurations with respect to $P_\cJ(s)$ are called thermal averages,
while the averages over samples, with respect to $\cP(\cJ)$, are
called quenched averages. I'll call $\cP(\cJ)$ the quenched
probability, to distinguish it from Boltzmann's probability.

Then one is lead to make a distinction between two
types of properties.

On the one hand, there are properties which
depend on the sample. For instance, the ground state configuration of
spins, the one which minimizes the energy, obviously depends on
$\cJ$. Actually, all the details of the energy landscape depend on $\cJ$.

On the other hand, some properties turn out to be 'self-averaging',
meaning that they are the same, for almost all samples (with a quenched
probability that goes to one in the large $N$ limit). For instance in
the EA or SK model the internal energy density
\begin{align}
U_\cJ=\frac{1}{N}\sum_s P_\cJ(s) E_\cJ(s)
  \end{align}
  is self averaging (this is easily proven in EA because one can cut a
  sample into many pieces and neglect the interactions between pieces
  which are of relative order surface to volume, it is less easy for
  the SK model \cite{SK}). This means that the distribution of $U_\cJ$ (when
  one picks a sample at random from the quenched probability) has a
  probability density that concentrates, when $N\to\infty$, around a
  given value $u$ that depends only on the inverse temperature $\beta$
  and on the statistical properties of the distribution of $\cJ$. The
  typical sample-to-sample fluctuations of $U_\cJ$ around this value are
  of order $1/\sqrt{N}$. In
  the limit $\beta\to \infty$, this also implies that the 
  ground state energy density is self-averaging. The same is true for
  all the extensive thermodynamic properties. For instance the
  magnetization density in presence of a magnetic field, or its linear
  dependence at small fields, the magnetic susceptibility, are
  self-averaging. This property of self-averageness is crucial: it is
  the reason why the measurements of magnetic susceptibilities or
  specific heat of two distinct
  spin glass samples with the same statistical properties (take for instance two
  sample of CuMn with 1$\%$ of Mn) give the same result: these are
  reproducible measurements because the measured property is
  self-averaging.

  Notice that, for the properties which are not self-averaging, one
  can study their quenched distribution. A typical example is the
  order parameter function that we shall discuss below\cite{MPSTV}.

\subsection{Second Challenge: Inhomogeneity}
  The second challenge that spin-glass theory had to face is
  inhomogeneity. The lesson we learn from detailed studies of the SK
  model is the following. For a typical sample $\cJ$, there exists a low-temperature 'spin-glass'
  phase in which the spins develop non-zero local magnetizations:
  \begin{align}
\langle s_i\rangle=m_i
  \end{align}
  Because of the disorder in the coupling constants $J_{ij}$, contrarily
  to the ferromagnetic case these
  magnetizations are not uniform. Analyzing a spin glass order in
  detail thus requires to use as order parameter the set of all the
  magnetizations. This is a $N$-component order parameter. Thouless Anderson and Palmer were 
  able to write a closed system of $N$ equations that relate all
  these components\cite{TAP}. The TAP equations, which generalize
  (\ref{meanfield}) to the spin-glass case, are:
  \begin{align}
m_i=\tanh\left[\beta\left(\sum_j J_{ij} m_j-\beta (1-q) m_i\right)\right]
    \end{align}
where $q=(1/N)\sum_j m_j^2$.
 With respect to the naive mean field equations $m_i=\tanh\left[\beta
   \sum_j J_{ij} m_j\right]$, they are characterized by the appearance
 of the ``Onsager reaction term''. This basically says that, when one
 computes the mean of the local magnetic field on site $i$, one should
 subtract from the naive estimate $ \sum_j  J_{ij} m_j $ the part of $m_i$
 which is polarized by $i$ itself. This means using a ``cavity''
 magnetization $m_j^c=m_j-\chi_j J_{ji}m_i$ where
 $\chi_j=\beta(1-m_j^2)$ is the local magnetic susceptibility of an
 Ising spin.

 When $N$ is not too large, say a few tens of thousands, TAP-like equations can
 be used as an algorithm, they can be solved by iteration using a
 specific iteration schedule that was found by Bolthausen\cite{bolthausen2014}. This
 gives information on the behavior of a given sample $\cJ$.

 On the other hand, when $N$ is very large, for instance of the order
 of the Avogadro number, one cannot write explicitely or solve the TAP
 equations. One must use a statistical study of the
 properties of these solutions. It turns out that this cannot be done
 directly on the TAP equations themsemeves, because the Onsager
 reaction term creates subtle correlations. The cavity method \cite{MPV_cavity,MPV} allows
 to circumvent this problem, by first analysing the statistics of the
 cavity field, the field acting on a spin in absence of this
 spin. This allows to build a full solution to the problem.

\subsection{Third Challenge: The many-valleys landscape}

Keeping to the SK model, it was found that there actually exist
  many different 'states' where the system can freeze, and therefore
  many solutions of the TAP equations. Each state
  $\alpha$ is characterized by $N$ magnetizations $m_i^\alpha$, so the
  order parameter is actually a $N$-component vector. This
  generalizes the situation of the ferromagnet. Instead of two states,
  identified by their average magnetization, we have many states. In
  each of them the average magnetization in absence of external field,
 $(1/N)\sum_i m_i^\alpha$, vanishes in the thermodynamic
 limit.

 Defining these states correctly is actually difficult. If one
 parallels the construction of the two pure states that we introduced
 in (\ref{ferro-mag}) for the ferromagnet, the natural generalization
 is to introduce for each state $\alpha$ a site-dependent small
 magnetic field $B_i^\alpha$ and take the limit where all these local
 fields go to zero after the thermodynamic limit. This leads to
 \begin{align}
m_i ^\alpha =\lim_{B^\alpha\to 0}\lim_{N\to\infty}
\langle s_i\rangle_{B^\alpha}
  \end{align}
 
The weakness of this definition is that we do not know how to choose
the local orientations of $B_i^\alpha$: on which site should they be
positive and on which site should they be negative? Solving this
problem requires knowing the signs of $m_i^\alpha$. So while this
definition of the order parameter is interesting, in practice it is
useless.

The replica method which was used to solve the SK model \cite{parisi_sol1,MPV} actually
has an interesting interpretation from this point of view. The idea is
that, if we do not know the preferred orientations where the spins
will polarize, the systems knows them. So, for theroretical
understanding, one can introduce, for a given sample $\cJ$, two
replicas of spins, $s$ and $\sigma$, with the same energy function
$E_\cJ$. In this noninteracting system, the probability of the two
configurations is
\begin{align}
  P_{\cJ}(s,\sigma)=\frac{1}{Z_\cJ^2}e^{-\beta (E_\cJ(s)+E_\cJ(\sigma))}
  \end{align}
  One can introduce the overlap $q=(1/N)\sum_i s_i \sigma_i$, and ask
  what is the distribution $P_\cJ(q)$ of this overlap, in the
  thermodynamic limit $N\to\infty$. In the high
  temperature paramagnetic phase, $P_\cJ(q)=\delta(q-q_0)$ where
  $q_0=0$ in absence of an external field, but it becomes $q_0>0$ in
  presence of a uniform field. In the spin glass phase, $P_\cJ(q)$
  becomes non trivial, it has a support $[q_0,q_1]$ with $q_1>q_0$,
  and it fluctuates from sample to sample. So this is a
  non-self-averaging quantity\cite{MPSTV}. Its quenched average, $P(q)=\int d\cJ
  \cP(\cJ) P_\cJ(q)$, is the order parameter for the spin glass
  phase. It is this order parameter which appears naturally, and is computed in the
  replica method with replica symmetry breaking, as shown in Parisi's
  seminal work\cite{Parisi83}.

  A simple way to
  define the existence of a spin glass phase using two replicas is to
  introduce a small coupling between them. The energy of a pair of
  configurations $s,\sigma$ now becomes:
  \begin{align}
E_\cJ^\epsilon(s,\sigma)=E_\cJ(s)+E_\cJ(\sigma)-\epsilon \sum_i s_i \sigma_i
  \end{align}
  Sampling the pairs of configurations with the corresponding
  Boltzmann weight, one can compute the expectation value of the
  overlap $\langle q\rangle^\epsilon=\int dq 
\;P_\cJ^\epsilon(q) q$. Taking the limit $\epsilon\to 0_\pm$ after
the thermodynamic limit, one finds
\begin{align}
q_1=\lim_{\epsilon \to 0^+}\lim_{N\to\infty} \langle q\rangle^\epsilon
  \ \ ;\ \
  q_0=\lim_{\epsilon \to 0^-}\lim_{N\to\infty} \langle q\rangle^\epsilon
  \end{align}
  These are the two limits of the support of $P(q)$. The existence of
  a spin glass phase is signalled by $q_1>q_0$. This definition gives
  a very intuitive interpretation to the use of replicas: one takes
  two replicas coupled by a small attractive interaction
  ($\epsilon>0$). When this interaction vanishes, if the spins in each
  of the two replicas
  remain correlated, this signals the spin glass phase. This criterion
  can also be used in glassy systems without disorder, like structural
  glasses \cite{franz_parisi,mm99}.

  The whole ``landscape structure'' of the spin glass phase can be
  analysed as follows: in a given sample there exist many pure states
  $\alpha$. Each of them is characterized by the $N$-dimensional vector
  of magnetizations $m^\alpha=\{m_1^\alpha,...,m_N^\alpha\}$, and its
  free energy $F^\alpha$. All the states that contribute to the
  thermodynamics have the same free-energy density $\lim_{N\to \infty}
  F^\alpha/N$, but they have finite free energy differences
  $F^\alpha-F^\gamma=O(1)$, and therefore each state contributes to the
  Boltzmann measure with a weight $P^\alpha$. Therefore
  \begin{align}
P(q)=\int d\cJ \cP(\cJ) \sum_{\alpha,\gamma}P^\alpha P^\gamma \delta\left(q-q^{\alpha\gamma}\right)
    \end{align}

    Various types of glassy phases are characterized by different
    types of $P(q)$ functions, two extremes being the simple ``one-step
    RSB''characteristic of the structural glass transition ones having
    only two $\delta$ peaks at $q_0$ and $q_1$ \cite{DerridaREM,GrossMezard,CrisantiSommers}, and the ``full-RSB''
    which occurs in the SK model and where the support of $P(q)$ is
    the full interval $[q_0,q_1]$, with an infinity of states
    organized in a hierarchical structure called ultrametric\cite{MPSTV},
    and a $\delta$ peak of $P(q)$ at the Edwards Anderson order parameter
    $q=q_1$, which gives the typical size of the states.

\subsection{Fourth Challenge: Out of equilibrium dynamics}
The last big challenge that spin glass theory had to face was the one
of equilibrium. The whole description that I gave so far is based on
the idea that a given sample of a spin glass can be characterized by
the Boltzmann measure $P_\cJ(s)$. However this is true only in the
case where the system reaches equilibrium. Experiments precisely
teach us that equilibrium is not reached in the spin glass phase. For
instance, measuring the magnetic susceptibility by first cooling the
system to a temperature $T<T_{sg}$ and then adding a small uniform
magnetic field $B$ gives a ``zero-field-cooled susceptibility'' $\chi_{ZFC}$ which
is different from the one found by placing the sample in the magnetic
field $B$ at high temperature (above the spin glass transition
$T_{sg}$), and then cooling it to $T$. This last procedure gives a
``field-cooled'' susceptibility $\chi_{FC}$ which is in general larger
than $\chi_{ZFC}$. In both cases the measurement of the susceptibility
is done at the same point $T,B$ of the phase diagram, but the results differ, proving that
the spin glass is out of equilibrium. Then a very legitimate question
is: how can the equilibrium theory be of any use ?

One of the first successes of the Parisi theory has been
to give a qualitative explanation of this difference by assuming that
the FC susceptibility corresponds to the reaction of the system when perturbing an equilibrium which
is a superposition of pure states, while the ZFC susceptibility
corresponds to a perturbation within one pure state (see eg
\cite{ParisiNobel}). In fact, if one
introduces a constrained perturbation to a SK spin glass, in which the
system reacts to a small magnetic field, but it is constrained to
remain at an overlap larger than $q$ from its initial state, then the
corresponding susceptibility is
\begin{align}
\chi(q)=\beta \int_q^1 dq'\; P(q') (1-q')
 \end{align}
 which gives in the two limiting cases:
 \begin{align}
\chi_{ZFC}=\beta \left[1-\int_0^1 dq'\; P(q') q'\right] \ \ ; \ \ \chi_{FC}=\beta \left[1-q_{1}\right]
 \end{align}
 
One can also go beyond, and try to study directly the dynamics of
mean-field models like the SK model. In the spin glass phase, the time
to reach equilibrium diverges in the thermodynamic limit. One can then
study what happens on various diverging time-scales, as in the first
works of Sompolinsky and Zippelius\cite{SompolinskyZippelius}.

An alternative approach which gives very interesting insight, is to solve the out of equilibrium
dynamics, as was proposed initially by Cugliandolo and Kurchan \cite{CugliandoloKurchan}. Focusing again on the mean-field models, one can
derive a closed set of equations for the two-time correlation
$C(t_w+t,t_w)=(1/N) \sum_i \langle s_i(tw+t)s_i(t_w)\rangle$ and the
two-time response function $R(t_w+t,t_w)$, which is
the linear response measured at time $t_w+t$ of a system which has started its dynamics at time
$0$, and to which a small magnetic field has been added at the time
$t_w$. In systems which reach their equilibrium, after a long waiting
time $t_w$, the
functions $C$ and $R$ become time-translation invariant, ie they
depend only on the measurement time $t$. This invariance is broken in the spin glass phase:
the $t$ dependance of these two functions depends on the age $t_w$ of
the system, and they keep evolving when $t_w$ increases, a phenomenon
called aging which is often observed in glassy systems. The simplest
scenario of aging would be one in which $C$ and $R$ become function
of $t/t_w^a$. For instance approximate $t/t_w$ scaling with $a=1$ is
often observed. In link with its hierarchical static structure, the SK model shows a more complicated 
behaviour, with various time-scales characterized by distinct
exponents $a$ playing a role.
In link with the aging phenomenon, one also finds a modification of
the standard fluctuation-dissipation theorem (FDT).

In an equilibrium system, at large enough $t_w$,  the
standard FDT relation between fluctuation and response is
$R(t)=\beta(C(0) - C(t))$ (notice that we use here an integrated
response function, as defined above).

In the spin glass phase, this is modified and becomes a relation between
$C(t_w,t)$ and $R(t_w,t)$ that holds
when both the waiting time $t_w$ and the measurement time $t$ are
large:
\begin{align}
\frac{\partial R(t_w,t)}{\partial t}=- \beta X(C(tw,t))\frac{\partial  C(t_w,t)}{\partial t}
\end{align}
The function $X(C)$ is the ``fluctuation dissipation ratio''. When
computed from spin-glass theory, one finds that it is equal to the
total probability of an overlap larger than $C$:
\begin{align}
X(C)=\int_C^1P(q) dq
  \end{align}
It can thus
be measured by plotting parametrically $R$ versus $C$. We have thus a
way to measure the equilibrium order parameter $P(q)$ from an out of
equilibrium measurement of correlation and response. This was done by
\cite{HerissonOcio} and the reader can find a discussion in \cite{ParisiNobel}.

\section{Statistical Physics of Inference}
\subsection{Machine learning as a statistical physics problem}
Spectacular recent developments of artificial intellignece are based
on machine learning. I'll sketch here the formal framework of
supervised learning, in order to relate it to statistical physics of
disordered systems. Recent introductions can be found in \cite{Zdeborova2015,Lauditi2023}.

Machine learning aims at learning a function from a $d$ dimensional
input $\xi \in \bbR^d$ to a $k$ dimensional output $y$. Usually one is
interested in large dimensional input like an image, so $d$ is large,
in practice it can be $10^6$ or more, and a small dimensional
output. Taking the famous example of handwritten digits, the image
could be an image of a digit, and the output would be the digit. In a
one-hot encoding, one would use $k=10$ and the digit $r$ would be
associated to $y_r=1$ and $y_{r'}=0$ for $r'\neq r$. One thus wants to
learn a target function $y=f_t(\xi)$. Actually in practical applications
we do not have a full definition of the function, but we have
examples, in the form of a database of pairs input-output
$\cD=\{\xi^\mu,y^\mu\}$, with $\mu \in \{1,...,P\}$.

\begin{figure}[t]
  \centering
  \includegraphics[width=.8 \textwidth]{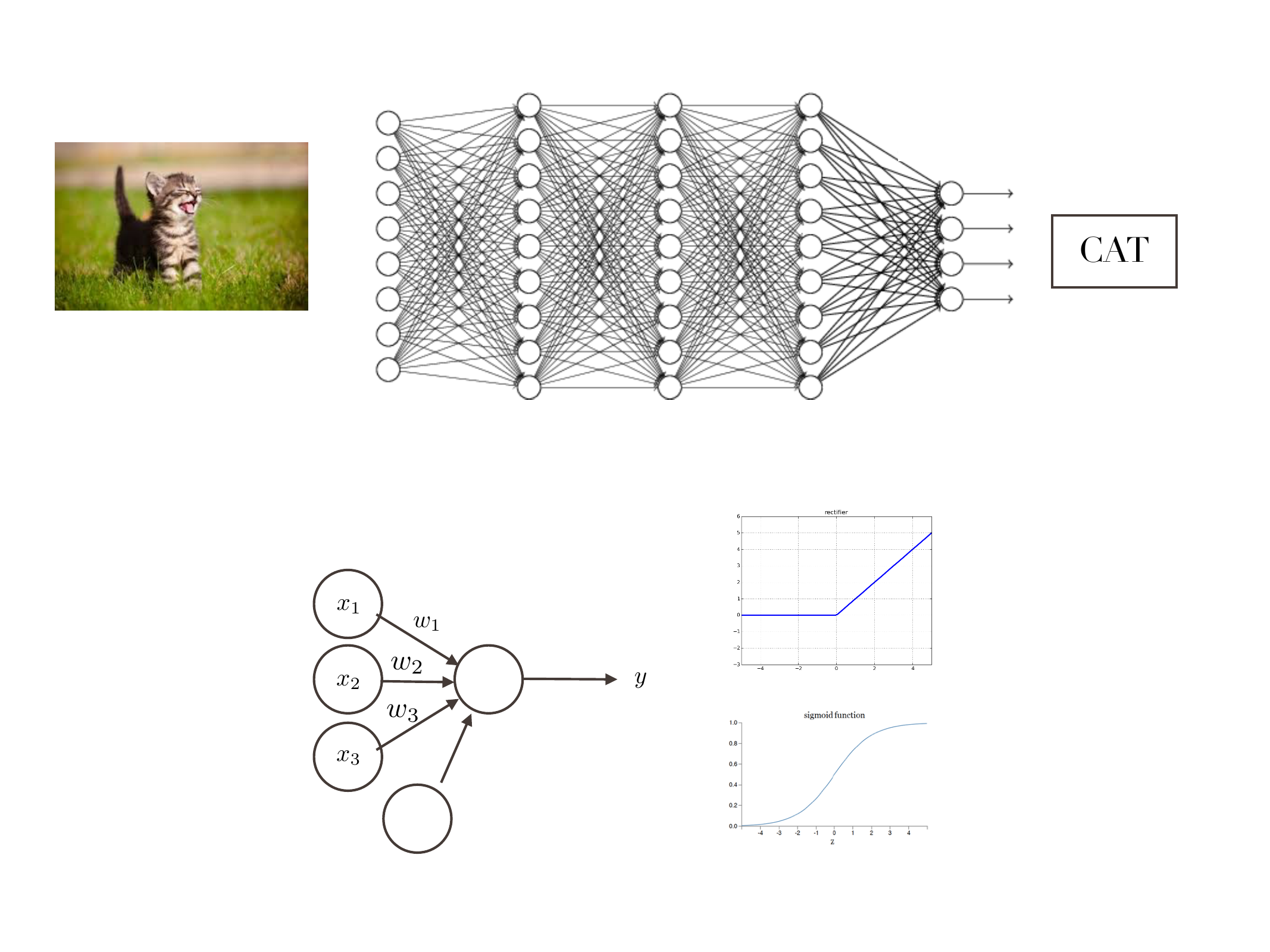}
\caption{\label{neurnet}\textbf{
Top: typical structure of a feedforward neurla network. The input
(image of a cat) is presented on the left layer. Data is processed, layer after
layer, until the output is given in the right. Bottom: Each layer is
built from artificial neurons. They receive inputs from the neurons of
the previous layer on their left hand side, these inputs are weighted,
and the linear combination of the reweighted inputs is then
transformed by a non-linear transfer function. Two examples of such
functions are shown here: the ReLu function and a sigmoid.
    }
  }
  \end{figure}

Modern deep networks are based on artificial neurons organized in
layers. Each neuron in layer $L$ is a simple unit that
receives a signal from the neurons in the previous layer, applies a
nonlinear function and sends this processed signal to the neurons of
the next layer (see Fig\ref{neurnet}). The activity of neuron $i$ in layer $r$ is given by
\begin{align}
x_i^r=\psi_i^r\left(\sum_j W^r_{ij} x_j^{r-1}\right)
\end{align}
where $W^r$ is a matrix of the ``synaptic efficacies'' between neurons
in layer $r-1$ and $r$. The nonlinear function $\psi_i^r$ can be for
instance a sigmoid
or a rectified linear unit. Usually it depends on the layer $r$ but
not on the precise neuron in the layer.

The layer $0$ is the input, $x^0=\xi$, and the layer $L$ is the
output, $x^L=y$; other layers are called hidden layers. A given realization of the neural network is given by
its architecture (the depth $L$ and the width of each layer), the
choice of nonlinear functions, and the values of the weights
$W=\{W^r\},\ r\in\{1,....,L\}$. We shall denote by $N$ the total
number of weights. If the width of each hidden layers is constant
equal to $h$, then $N=d h+(L-2) h^2+ h k$. A given network with parameters $W$
implements a function
from input to ouptut $y=f(W,\xi)$.

In most applications the architecture
is chosen by the engineer, based on previous experience, but the
weights are learnt. Indeed, machine learning designates the process by which the parameters
(in this case the weights) are not given to the machine, but the
machine learns them from data. This learning is called the training
phase. In order to train a network one defines a ``loss function''
$L(W)$ which measures the errors made by the machine with parameters
$W$ on the database. For instance one could use a quadratic loss
\begin{align}
  L_\cD(W)=\sum_{\mu=1}^P \left(y^\mu-f(W,\xi^\mu)\right)^2
  \label{loss-def}
  \end{align}
but many other choices are possible. Then the training phase consists
in finding the $W$ that minimize the loss. In practice, people use a
form of gradient descent called stochastic gradeint descent, in
which one moves in the $W$ landscape using iteratively noisy versions of the
gradient computed from partial sums involving some batches of $\mu$
indices.

Once the learning has been done, one can use the couplings $W^*$ which
have been found during training, and see how well the network
generalize when it is presented some new data that it has never
seen. The test -or generalization- loss has the same expression as
(\ref{loss-def}), but with new, previously unseen, input-output pairs.

\subsection{Data as disorder}
One can also introduce a probability distribution in the space of
weights $W$, of the form
\begin{align}
  P_\cD(W)=\frac{1}{Z_\cD}P_0(W) e^{-\beta L_\cD(W)}
  \label{PdeW-def}
  \end{align}
where $P_0(W)$ is a prior on the weights, one can choose it as a
factorized prior $P_0(W)=\prod_{r,i,j}\rho(W_{ij}^r)$ (for instance
one can use for $\rho$ a
gaussian if one wants to avoid too large weights). One could also normalize
$\sum_j (W_{ij}^r)^2=1$, but I will keep here for simplicity to the
factorized case. The parameter  $\beta$
is an auxiliary inverse temperature parameter. When $\beta$ is large
this probability distribution is concentrated on the sets of weights
$W$ which minimize the loss. This formalism amounts to an approach of
statistical physics in the space of weights. It was pioneered by
Elizabeth Gardner \cite{Gardner,GardnerDerrida} in the study of the simplest network, the
perceptron which has no hidden unit (and is therefore limited to
linearly separable tasks). 

In recent applications of deep networks like the large language model
Chat-GPT, the total number of weights can be of order $10^{11}$, and
the total number of operations used in order to train a network on
extremely large databases of basically all available text is easy to
remember, it is of the order of $10^{24}$, a ``mole of operations''. So
the distribution (\ref{PdeW-def}) is a measure in a large $N$-dimensional
space. The elementary variables, the weights, are real valued
variables with a measure $\rho$. They are coupled through an energy
which is the loss  $L_\cD(W)$. This energy is a complicated function
of the variables $W$, and it depends on a large set of parameters, namely
all the input-output pairs $\cD$. Therefore the measure
(\ref{PdeW-def}) has all the ingredients of a statistical physics
systems with a quenched disorder which is the data.

Having in mind our discussion of disordered systems, we can
immediately identify several
questions. One can work on a given data base (a given sample) and ask
about the landscape for learning, and for generalization. But clearly,
for theoretical studies, one would like to have an ensemble of
samples, which is a probability measure in the space of inputs, and
for each input the corresponding output. With such a setup of a {\it
  {data ensemble}}, one can draw a database by choosing inputs
independently at random from the ensemble, one can study the property
of self-averageness (which properties of the optimal network $W^*$ are
dependent on the precise realization of the database, and which ones
are not - in the large $N$ limit ?). The test loss becomes easy to
define: it is the expectation value of the loss, over pairs of
input-output generated from the data ensemble.

Working with a single dataset or with a data ensemble are two rather
different approaches. In many practical applications the engineer's
approach is to use a single dataset,
and the definition of an ensemble is not obvious. For instance if one
wants to identify if there is a cat or a dog on an image, so far what
is done is use huge databases of images of cats and dogs, randomly
chosing part of them
for training, and another part for the test phase. However,
in such a single database setup it is not easy to develop a theory: on
the one hand, there is a risk of developing a theory which is too much
tailored to this precise database, and from which one cannot draw
general  conclusions; also, one cannot use probabilities to compute
the generalization error. So the use of data ensembles is
clearly welcome from a theoretical point of view, but then one 
faces the difficulty of defining the ensemble in such a way that it
will include some essential features of real databases, but it should
be smooth enough so that one can interpolate through it reasonably (the ensemble which would
use a probability law that is a sum of $\delta$ peaks on each point of
a database
is useless), and simple enough that it can be studied. So the question
of finding good ensembles is a fundamental question of how to model
the ``world'', ie the set of all possible inputs that can be
presented. Here by modelling one intends it in a physics' approach,
namely being able to identify the key features that should be
incorporated into the ensemble, neglecting less important ``details''.

Interestingly, this quest for modeling of data meets with an
important recent direction of development of machine learning, which are generative
models. In parallel to, and in symbiosis with the supervised machine
learning that I have briefly exposed, very significant progress has
been made on generative models. These aim at generating data 'similar
to' a given database. Among the processes that have been explored, one
can mention Generative Adversarial Networks (GAN), or the very
physical generative diffusion models which take a database, degrade it
using a Langevin process until it has been transformed into
pure noise, and then reverse the Langevin process to reconstruct
artificial data from noise (see \cite{Sohl_Dickstein2015,
Song2019,Song_Sohl-Dickstein2021}); for a recent review see \cite{yang2022diffusion}, and
for a statistical
physics perspective: \cite{BiroliMezard23_GenDiff}).

\subsection{Surprises}
\subsubsection{Perceptrons}
Training a neural network in supervised learning amounts to
finding the ground state of a strongly disordered system. One can thus
ask what properties of spin glasses one can find in neural
networks. Early studies on perceptrons have provided important
benchmarks. Two main categories of tasks have been studied: learning
arbitrary labels, or learning from a ``teacher rule''. In both cases,
the database consists of 
$P$ independent input datapoints
each consisting of  $d$ iid numbers from a distribution $\rho_0(\xi)$.
In the case of
learning from arbitrary labels, for each input one generates a desired output which is drawn
randomly, independently from the input. In this case one studies only
the training phase, generalization has no meaning. In the case of
learning a teacher
rule, one generates the output from a ``teacher'' set of weights, $W_t$,
through $y=f(W_t,\xi)$. The quality of training can be monitored by
computing  some distance between $W^*$ and $W_t$, like for instance $|W^*-W_t|^2/|W_t^2|$.

The behaviour of the training depends a lot on the a priori measure on the
weights, $\rho$. If $\rho$ is gaussian, or imposes a spherical
constraint, the training problem is convex and the landscape is
simple. The training and generalization error decrease continuously
with $\alpha=P/N$. If $\rho$ is discrete, corresponding to Ising
spins, then the training on random labels shows a replica symmetry
breaking phase at $\alpha_c=.83$ \cite{KrauthMezard} which has has a
strange nature. On the one hand, the typical configurations at
$\alpha$ close to $\alpha_c$ are isolated points, building a
golf-course potential \cite{huang2013entropy}. On the other hand, atypical,
exponentially rare regions of phase space concentrate a large number
of neighboring solutions\cite{baldassi2015subdominant}, and are easy
to find.  As far as learning a teacher rule is concerned, with
binary synapses, the
generalization error shows a first order phase transition to perfect
generalization\cite{gyorgyi1990first} when $\alpha$ is larger than a threshold
$\alpha_g\simeq 1.25$. The phase diagram can be studied rigorously, and
when $\alpha>\simeq 1.5$ one can also use iterative message-passing
algorithms based on the cavity-TAP method \cite{Mezard1989} in order to find the optimal weights defined by the teacher\cite{barbier2017phase}.

\subsubsection{Deep networks}

The recent experimental successes of deep networks have triggered a
lot of analyses, but the situation is less clear than in
perceptrons. So far, statistical physics approaches can be used efficiently in
multilayer networks either when the transfer functions are linear \cite{LiSompolinsky}, or when
there is a single layer of learnable parameters with a size that
diverges in the thermodynamic limit, like for instance in committee
machines or parity machines.

Also, the empirical observations of the
learning process show a picture which is rather different from the
usual spin glass landscape. The first obsevation is that very complex
functions involving billions of weights are learnable from
examples, using the simple stochastic gradient descent algorithm. This means that the loss
function which is optimized is not as rough as one would have in a
spin glass. Typically, stochastic gradient descent, when initiated from generic
initial conditions (with small weights), finds a set of weights $W^*$
not far from the initial condition, which has a 
small loss. Surprisingly, in this large dimensional space, 
the set of weights with  small is not sparse, as one could have expected
from our experience of optimizing in large dimensions.

Once the network has been trained, typically using a number $N$ of
weights that is of the same order as the number of points in the
database, one must study its generalization properties. From a
statistics' perspective, what has been done in the training phase is
fitting a complicated $N$ dimensional function using $P$
datapoints. This is possible because $N$ is large, but one should
expect to be in a regime of overfitting, and therefore a poor
generalization. This is not the case. Actually, increasing the depth
of the network, and therefore the number $N$ of fitting parameters,
one observes that the generalization errors keeps decreasing, while it should
shoot-up in the overfitting regime.
Among all these minima of the loss, some generalize better than
others, and this seems to be correlated with the flatness of the
landscape around the minimum\cite{baldassi2020shaping}.

These facts indicate that deep-learning landscapes are
rather different from the ones that have been explored in spin glass
theory or in perceptrons. What are the ingredients responsible for
the relatively easy training and the lack of overfitting in deep
networks? Three directions are being
explored: 1) the architecture of the networks, and in particular the
importance of using deep enough networks, with many layers; in
practice the design of the architecture, including the choice of
nonlinearities, is an engineer's decision based on previous
experience; 2) the learning algorithm; stochastic gradient descent
started from weights with small values seems efficient at finding out
first the main pair correlation in the data, then gradually
improving \cite{refinetti2023}; 3) The structure of data: practical problems deal with
highly structured data, whether they are text, image, amino-accid
sequences. In the next section I shall  focus on this last point,  argue about the relevance of
structured data and describe the challenge it poses to statistical physics. 

\section{The new challenge of spin-glass theory: structured disorder}
Data is highly structured, and a major objective is to develop mathematical models for the datasets on which neural networks
are trained. 
Most theoretical results on neural networks do not model the structure
of the training data. Statistical learning theory~\cite{Vapnik2013,
  Mohri2012} usually provides bounds that hold in the worst case, but
are far from describing typical properties seen in experiments. On the
other hand, traditional statistical physics approaches use a setup
where inputs are either drawn
component-wise i.i.d.\ from some probability distribution, or are
gaussian distributed~\cite{Seung1992, Engel2001}.  Labels are either
random or given by some random, but fixed function of the input. Despite
providing valuable insights, these approaches ignore key
structural properties of real-world datasets.

In recent years several aspects of data structure have been explored,
and the first 
ensembles of structured data have started to be
developed. The challenge is of course to create ensembles which
contain some of the essential structure, but are at the same time simple enough to
be analysed. I will mention here three categories of data properties which are being studied:
effective dimensionality, correlations, and combinatorial/hierarchical structure.

\subsection{Effective dimension}
 Let us consider perhaps the simplest canonical problem of
supervised machine learning: classifying the handwritten digits in the MNIST
database using a neural network~\cite{lecun1998}. The input
patterns are images with $28\times 28$ pixels, so \emph{a priori} we work in the
high-dimensional space $\mathbb{R}^{784}$. However, the inputs that may be interpreted
as handwritten digits, and hence constitute the ``world'' of our problem, span
but a lower-dimensional manifold within $\mathbb{R}^{784}$. Although
this manifold is not easily defined,  its dimension can be estimated 
based on the distance between neighboring points in the dataset~\cite{Grassberger1983,
  Costa2004, Levina2004a, Spigler2019}. In fact, if we consider $P$
independent datapoints in a $D$ dimensional space, we expect that the
distance between nearest neighbors scales like $P^{-1/D}$. Analysing the
MNIST data base, one finds the effective dimension
to be around $D\approx 15$, much smaller than $N= 784$. The
``perceptual submanifold'' associated with each digit also has an
effective dimension, ranging from $ \approx 7 $ for the digit 1 to
$\approx 13$ for the digit 8 \cite{HeinAudibert}. Therefore
the task of identifying a handwritten digit consists in finding
these ten perceptual submanifolds, embedded in the 15-dimensional
``world'' manifold of handwritten digits. Of course, the
problem is that these manifolds are nonlinear, folded, and it is
hard to find them (see \cite{fefferman} for algorithmic
approaches). The same phenomenon of reduction in effective dimension is
found in other datasets. For instance, images in CIFAR10 are defined
in dimension $N=1024$, but have an effective dimension $D\approx 35$.
In most machine learning problems, the effective ``world'' on
which we train our networks has an effective dimension $D\ll N$ (in
fact, a good practice would be to
train the networks so that they can identify when they see  an input
which is far from
the world in which they were trained, and refuse to give an answer in
such cases).

A simple attempt at including this effective dimensionnality in
ensemble of data is the ``hidden manifold model'' \cite{goldt2020}. 
In this model, the seed $s$ of a datapoint is generated iid in a
$D$-dimensional ``latent'' space, for instance from a gaussian
distribution. Then the datapoint components $\xi_i$ are generated as
\begin{align}
  \xi_i= g\left(\sum_{r=1}^D F_{ir}s_r\right)
  \label{hidden-manifold}
\end{align}
where $F_{ir}$ are given and define the model, as well as $g$ which is
a nonlinear function. It turns out that, when the components of $F$
are well balanced (and in particular if they are generated iid from a
well behaved distribution), one can generalize the
statistical physics studies of the perceptrons or shallow networks to
data which has this hidden manifold structure. The reason is that the
inputs of hidden units actually receive an input which becomes
gaussian distributed. This ``gaussian equivalence theorem'' allows to use the whole traditional
spin-glass machinery. It also tells that this kind of model has its
limitations, as it is equivalent to some type of gaussian distributed
inputs. 

Note that the hidden manifold structure of data defined in
(\ref{hidden-manifold}) can receive a different interpretation, where
one would like to learn from the latent signal $s$ in $D$ dimensions,
but one first projects it to a $N$-dimensional space of random features
which are fixed, and not learnt, a problem which has been studied in
detail when the matrix $F$ is generated from a random matrix ensemble\cite{mei2022generalization}
(but gaussian equivalence holds beyond this, as long as matrix
elements are well balanced, like for instance in Hadamard
transformation).

Actually the construction of hidden manifolds can be elaborated by
using, instead of (\ref{hidden-manifold}), an iterative construction
based on several layers of projections, as done in the GAN
approach. In that case gaussian equivalence is conjectured to hold,
although it has not been proven yet \cite{goldt2020}.

\subsection{Correlations}
From the database, one can construct the empirical pair correlation
$C_{ij}$ between two components of the input
\begin{align}
C_{ij}=\frac{1}{P}\sum_{\mu=1}^P \xi_i^\mu \xi_j^\mu
\end{align}
as well as higher order correlations
(here we assume that we use centered data, in which the empirical mean
of $\xi_i$ has been substracted). A distinguishing property of
practical datasets is that correlations are highly structured, and
actually some of this structure is already seen at the level of the
pair correlation.

For instance, if one diagonalizes the matrix of pair correlations,
which is of Wishart type, one  finds a spectrum of eigenvalues which
differs notably from the Marcenko Pastur one that would be obtained
if the components $\xi_i^\mu$ were distributed independently and
identically.
Instead, one typically gets  a power
law distribution of the large eigenvalues, and it has
been argued that this power-law scaling is actually related to the
power-law decay of the loss with respect to either $N$ or $P$, found in
large language models \cite{maloney2022solvable}.
A simple attempt at including this effective dimensionnality in an
ensemble of data is to use random Wishart matrices with a power-law
distributed spectrum\cite{maloney2022solvable,LeviOz}.

Note that this power-law scaling (with small exponents) of eigenvalues
of the correlation matrix points  to the existence of some type of
long-range correlations. In fact very structured and long-range
correlations in data are very important, and the recently developed
''attention mechanism''  is precisely built in order to handle such
correlations\cite{vaswani2017attention}. These are of a type which is rather different from what
one is used to in statistical physics. The easiest way to illustrate
them is through language models. In these models, one decomposes
the sentences into tokens (typically words or -for composite words-
portions of words) and the language models are trained from a large
corpus, at the task which is to take a text, interrupt it somewhere,
and give the best guess of the next token. Clearly the simplest
approach would be to sample the conditional probability distribution :
take the previous $k$ tokens before being
interrupted, and look in the database at sentences which have exactly
this sequence of k tokens, and compute from this database the most
probable next token. This approach was started very early on, by
Shannon himself. But clearly it is limited to small values of
$k$: beyond $k$ of order a couple of dozens, one des not have the
statistics to infer the conditional probability. But it turns out that
key tokens, which are crucial for guessing the next one,  can be found
much earlier in the text. Take for instance this sentence written above
: ``{\it Instead, one gets typically a power
law distribution of the large eigenvalues, and it has
been argued that this power-law scaling is actually related to the
power-law decay of the}. In order to guess the next word, 'loss', it
would be useful to focus on portions of this paper which appear much
earlier, where the loss is defined. It is this type of long-range
correlation that is handled by the attention mechanism.

\subsection{Combinatorial and hierarchical structure}
A third distinctive structure of datasets used in practice is its
combinatorial nature. Imagine for instance a photo of a lecture
hall: it is composed by a group of students, each sitting at his
desk. Then each student is ``composed'' of head, chest, arms, and each
head is ``composed'' of eyes, nose, mouth, hair, and the eyes are
``composed'' by  pigmented epithemial cells, etc. This is actually
typical, and most of the images that we want to analyse have this type
of combinatorial structure with a hierarchy of features and subfeatures related to
the scale at which one looks. This stucture is also related to the
decoding that happens when learning from images with a deep network:
one typically finds that the first layers of the network decode small
scales elements like edges, and going further into the network one
gradually 
identifies  larger scale  properties, until in the final layers one is
able to decide the content of an image. Interestingly, the same type of analysis, from small
scale to larger scales, takes place in the sequence of visual areas used
in the brains of primates. One also finds the same combinatorial/hierarchical
structure in text for instance, and also in protein sequences with
their primary, secondary and ternary structures.

The first attempts at building ensembles with
combinatorial/hierarchical properties are still rather rudimentary. An
easy case, although not very realistic, is the one of linear
structures. Interestingly, one can show that an associative memory
network\cite{Hopfield82} trying to store such hierarchical patterns can
be mapped onto a layered network where the first layers analyze the
small scale features, and the information is then built gradually to
larger scales, by combining smaller scale features of previous
layers.   Very recently, simple nonlinear versions of
combinatorial/hierarchical data ensembles have started to be
explored\cite{PetriniHierarchical,mossel2016deep}.

\section{Conclusion}
Constructing a theory of deep learning is an important challenge, both
from the theoretical point of view, but also for applications: only a
solid theory will be able to turn a deep network prediction from a
black-box best guess into a statement which can be explained and justified,
and whose worst-case behaviour can be controlled. The main high-level
challenge that is faced in deep network is the one of emergence: how
is the information gradually elaborated when it is processed from
layer to layer in the network? How is it encoded collectively?
Contemporary networks are working in a high dimensional regime, and
what we need is a good control of the representations obtained  from
data of probability distributions in large dimensions. This is
typically a problem of statistical physics. One big question is
whether we will be able to elaborate a statistical physics of deep
network which is based on a not-too-large number of order parameters,
that can be controlled statistically, as was done in spin glasses.

In
order to be relevant, this appraoch to deep networks must be able to
take into account important ingredients of the reald 'world', and in
particular its structure. So far spin glass theory has been developed
mostly for ensembles in which the coupling constants are identically
and independently distributed. It is known that more stuctured
ensembles can be very hard to study. This is the case for instance of
the EA model: in this model,  the fact that the spins are coupled only among nearest
neighbours on a cubic lattice is a type of Euclidean structure, and
this problem has not been solved exactly so far. A fascinating new
challenge of spin glass theory is to develop new ensembles of
correlated disorder, including 
some of the most relevant ingredients that are found in real databases,
like long-range correlations, hierarchy, combinatorial structures, effective
dimensions, while being able to keep some analytic control of the
problem.

\bibliographystyle{plain}
\bibliography{refs_all.bib}

\begin{thebibliography}{10}

\bibitem{andersonMore}
PW~Anderson.
\newblock More is different.
\newblock {\em Science}, 177:393--396, 1972.

\bibitem{baldassi2015subdominant}
Carlo Baldassi, Alessandro Ingrosso, Carlo Lucibello, Luca Saglietti, and
  Riccardo Zecchina.
\newblock Subdominant dense clusters allow for simple learning and high
  computational performance in neural networks with discrete synapses.
\newblock {\em Physical review letters}, 115(12):128101, 2015.

\bibitem{baldassi2020shaping}
Carlo Baldassi, Fabrizio Pittorino, and Riccardo Zecchina.
\newblock Shaping the learning landscape in neural networks around wide flat
  minima.
\newblock {\em Proceedings of the National Academy of Sciences},
  117(1):161--170, 2020.

\bibitem{barbier2017phase}
Jean Barbier, Florent Krzakala, Nicolas Macris, L{\'e}o Miolane, and Lenka
  Zdeborov{\'a}.
\newblock Phase transitions, optimal errors and optimality of message-passing
  in generalized linear models, 2017.

\bibitem{BiroliMezard23_GenDiff}
G.~Biroli and M.~M\'ezard.
\newblock Generative diffusion in very large dimensions.
\newblock {\em arxiv preprint arXiv:2306.03518}, 2023.

\bibitem{bolthausen2014}
Erwin Bolthausen.
\newblock An iterative construction of solutions of the tap equations for the
  sherrington--kirkpatrick model.
\newblock {\em Communications in Mathematical Physics}, 325(1):333--366, 2014.

\bibitem{FarBeyond}
P.~Charbonneau, M.~M{\'e}zard, E.~Marinari, G.~Parisi, F.~Ricci-Tersenghi,
  G.~Sicuro, and F.~Zamponi, editors.
\newblock {\em Spin-Glass Theory and Far Beyond}.
\newblock World Scientific, Singapore, 2023.

\bibitem{Costa2004}
J.A. {Costa} and A.O. {Hero}.
\newblock Learning intrinsic dimension and intrinsic entropy of
  high-dimensional datasets.
\newblock In {\em 2004 12th European Signal Processing Conference}, pages
  369--372, 2004.

\bibitem{CrisantiSommers}
A.~Crisanti and H.J. Sommers.
\newblock The spherical pbspin interaction spin glass model the statics.
\newblock {\em Zeitschrift für Physik B Condensed Matter}, 87:341--354, 1992.

\bibitem{CugliandoloKurchan}
L.~Cugliandolo and J.~Kurchan.
\newblock Analytical solution of the off equilibrium dynamics of a long range
  spin-glass model.
\newblock {\em Phys.Rev.lett.}, 71:173, 1993.

\bibitem{DerridaREM}
Bernard Derrida.
\newblock {Random-energy model: An exactly solvable model of disordered
  systems}.
\newblock {\em Physical Review B}, 24(5):2613--2626, sep 1981.

\bibitem{EA}
S.~F. Edwards and P.~W. Anderson.
\newblock Theory of spin glasses.
\newblock {\em J. of Physics F}, 5:965, 1975.

\bibitem{Engel2001}
A.~Engel and C.~{Van den Broeck}.
\newblock {\em {Statistical Mechanics of Learning}}.
\newblock Cambridge University Press, 2001.

\bibitem{fefferman}
Charles Fefferman, Sanjoy Mitter, and Hariharan Narayanan.
\newblock Testing the manifold hypothesis.
\newblock {\em Journal of the American Mathematical Society}, 29:983--1049,
  2016.

\bibitem{franz_parisi}
Silvio Franz and Giorgio Parisi.
\newblock Recipes for metastable states in spin glasses.
\newblock {\em Journal de Physique I}, 5(11):1401--1415, 1995.

\bibitem{Gardner}
E.~Gardner.
\newblock The space of interactions in neural network models.
\newblock {\em Journal of Physics A: Mathematical and General}, 21:257, 1988.

\bibitem{GardnerDerrida}
Elizabeth Gardner and Bernard Derrida.
\newblock Three unfinished works on the optimal storage capacity of networks.
\newblock {\em Journal of Physics A: Mathematical and General}, 22(12):1983,
  1989.

\bibitem{goldt2020}
Sebastian Goldt, Marc M{\'e}zard, Florent Krzakala, and Lenka Zdeborov{\'a}.
\newblock Modeling the influence of data structure on learning in neural
  networks: The hidden manifold model.
\newblock {\em Physical Review X}, 10(4):041044, 2020.

\bibitem{Grassberger1983}
Peter Grassberger and Itamar Procaccia.
\newblock Characterization of strange attractors.
\newblock {\em Physical review letters}, 50(5):346, 1983.

\bibitem{GrossMezard}
D.~J. Gross and M.~M\'ezard.
\newblock The simplest spin glass.
\newblock {\em Nucl. Phys. B}, 240:431--452, 1984.

\bibitem{gyorgyi1990first}
G{\'e}za Gy{\"o}rgyi.
\newblock First-order transition to perfect generalization in a neural network
  with binary synapses.
\newblock {\em Physical Review A}, 41(12):7097, 1990.

\bibitem{HeinAudibert}
Matthias Hein and Jean-Yves Audibert.
\newblock Intrinsic dimensionality estimation of submanifolds in rd.
\newblock In {\em Proceedings of the 22nd international conference on Machine
  learning}, pages 289--296, 2005.

\bibitem{HerissonOcio}
Didier H{\'e}risson and Miguel Ocio.
\newblock Fluctuation-dissipation ratio of a spin glass in the aging regime.
\newblock {\em Physical Review Letters}, 88(25):257202, 2002.

\bibitem{Hopfield82}
J~J Hopfield.
\newblock Neural networks and physical systems with emergent collective
  computational abilities.
\newblock {\em PNAS}, 79(8):2554--2558, 1982.

\bibitem{huang2013entropy}
Haiping Huang, KY~Michael Wong, and Yoshiyuki Kabashima.
\newblock Entropy landscape of solutions in the binary perceptron problem.
\newblock {\em Journal of Physics A: Mathematical and Theoretical},
  46(37):375002, 2013.

\bibitem{KrauthMezard}
Werner Krauth and Marc M{\'e}zard.
\newblock Storage capacity of memory networks with binary couplings.
\newblock {\em Journal de Physique}, 50(20):3057--3066, 1989.

\bibitem{Lauditi2023}
C.~Lauditi, E.~Troiani, and M.~M\'ezard.
\newblock Sparse representations, inference and learning.
\newblock {\em arXiv preprint arXiv:2306.16097}, 2023.

\bibitem{lecun1998}
Y.~LeCun and C.~Cortes.
\newblock {The MNIST database of handwritten digits}, 1998.

\bibitem{LeviOz}
Y.~Levi, N. andd~Oz.
\newblock The underlying scaling laws and universal statistical structure of
  complex datasets.
\newblock {\em arXiv:2306.14975}, 2023.

\bibitem{Levina2004a}
E.~Levina and P.J. Bickel.
\newblock {Maximum likelihood estimation of intrinsic dimension}.
\newblock In {\em Advances in Neural Information Processing Systems 17}, 2004.

\bibitem{LiSompolinsky}
Qianyi Li and Haim Sompolinsky.
\newblock Statistical mechanics of deep linear neural networks: The
  backpropagating kernel renormalization.
\newblock {\em Physical Review X}, 11(3):031059, 2021.

\bibitem{maloney2022solvable}
Alexander Maloney, Daniel~A Roberts, and James Sully.
\newblock A solvable model of neural scaling laws.
\newblock {\em arXiv preprint arXiv:2210.16859}, 2022.

\bibitem{mei2022generalization}
Song Mei and Andrea Montanari.
\newblock The generalization error of random features regression: Precise
  asymptotics and the double descent curve.
\newblock {\em Communications on Pure and Applied Mathematics}, 75(4):667--766,
  2022.

\bibitem{Mezard1989}
M~M{\'e}zard.
\newblock The space of interactions in neural networks: Gardner's computation
  with the cavity method.
\newblock {\em Journal of Physics A: Mathematical and General}, 22(12):2181,
  1989.

\bibitem{mm99}
M.~M\'ezard.
\newblock How to compute the thermodynamics of a glass using a cloned liquid.
\newblock {\em Physica A}, 265:352--367, 1999.

\bibitem{MPSTV}
M.~M\'ezard, G.~Parisi, N.~Sourlas, G.~Toulouse, and M.A. Virasoro.
\newblock On the nature of the spin glass phase.
\newblock {\em Phys. Rev. Lett.}, 52:1156, 1984.

\bibitem{MPV}
M.~M{\'e}zard, G.~Parisi, and M.~A. Virasoro.
\newblock {\em Spin-Glass Theory and Beyond}.
\newblock World Scientific, Singapore, 1987.

\bibitem{MPV_cavity}
M~M{\'e}zard, G~Parisi, and MA~Virasoro.
\newblock Sk model: The replica solution without replicas.
\newblock {\em Europhys. Lett}, 1(2):77--82, 1986.

\bibitem{Mohri2012}
M.~Mohri, A.~Rostamizadeh, and A.~Talwalkar.
\newblock {\em {Foundations of Machine Learning}}.
\newblock MIT Press, 2012.

\bibitem{mossel2016deep}
Elchanan Mossel.
\newblock Deep learning and hierarchal generative models.
\newblock {\em arXiv preprint arXiv:1612.09057}, 2016.

\bibitem{parisi_sol1}
G~Parisi.
\newblock Infinite number of order parameters for spin glasses.
\newblock {\em Physical Review Letters}, 43(23):1754, 1979.

\bibitem{Parisi83}
G.~Parisi.
\newblock Order parameter for spin glasses.
\newblock {\em Physical Review Letters}, 50:1946, 1983.

\bibitem{ParisiNobel}
G.~Parisi.
\newblock Nobel lecture: Multiple equilibria.
\newblock {\em arXiv preprint arXiv:2304.00580}, 2023.

\bibitem{PetriniHierarchical}
L.~Petrini, F.~Cagnetta, M.~Tomasini, A.~Favero, and M.~Wyart.
\newblock How deep neural networks learn compositional data: The random
  hierarchy model.
\newblock {\em arXiv:2307.02129}, 2023.

\bibitem{refinetti2023}
M.~Refinetti, A.~Ingrosso, and S.~Goldt.
\newblock Neural networks trained with sgd learn distributions of increasing
  complexity.
\newblock In {\em ICML 2023}, 2023.

\bibitem{Seung1992}
H.~S. Seung, H.~Sompolinsky, and N.~Tishby.
\newblock {Statistical mechanics of learning from examples}.
\newblock {\em Physical Review A}, 45(8):6056--6091, 1992.

\bibitem{SK}
David Sherrington and Scott Kirkpatrick.
\newblock Solvable model of a spin glass.
\newblock {\em Physical Review letters}, 35(26):1792, 1975.

\bibitem{Sohl_Dickstein2015}
Jascha Sohl-Dickstein, Eric Weiss, Niru Maheswaranathan, and Surya Ganguli.
\newblock Deep unsupervised learning using nonequilibrium thermodynamics.
\newblock In {\em International Conference on Machine Learning}, 2015.

\bibitem{SompolinskyZippelius}
H.~Sompolinsky and A.~Zippelius.
\newblock Dynamic theory of the spin glass phase.
\newblock {\em Phys.Rev.Lett.}, 47:359, 1981.

\bibitem{Song2019}
Yang Song and Stefano Ermon.
\newblock Generative modeling by estimating gradients of the data distribution.
\newblock {\em Advances in Neural Information Processing Systems}, 2019.

\bibitem{Song_Sohl-Dickstein2021}
Yang Song, Jascha Sohl-Dickstein, Diederik~P Kingma, Abhishek Kumar, Stefano
  Ermon, and Ben Poole.
\newblock Score-based generative modeling through stochastic differential
  equations.
\newblock In {\em International Conference on Learning Representations}, 2021.

\bibitem{Spigler2019}
S.~Spigler, M.~Geiger, and M.~Wyart.
\newblock {Asymptotic learning curves of kernel methods: empirical data v.s.
  Teacher-Student paradigm}.
\newblock {\em arXiv:1905.10843}, 2019.

\bibitem{TAP}
David~J. Thouless, Philip~W. Anderson, and Robert~G. Palmer.
\newblock Solution of 'a solvable model of a spin glass'.
\newblock {\em Philosophical Magazine}, 35(3):593--601, 1977.

\bibitem{Vapnik2013}
V.~Vapnik.
\newblock {\em {The nature of statistical learning theory}}.
\newblock Springer science {\&} business media, 2013.

\bibitem{vaswani2017attention}
Ashish Vaswani, Noam Shazeer, Niki Parmar, Jakob Uszkoreit, Llion Jones,
  Aidan~N Gomez, {\L}ukasz Kaiser, and Illia Polosukhin.
\newblock Attention is all you need.
\newblock {\em Advances in neural information processing systems}, 30, 2017.

\bibitem{yang2022diffusion}
Ling Yang, Zhilong Zhang, Yang Song, Shenda Hong, Runsheng Xu, Yue Zhao,
  Yingxia Shao, Wentao Zhang, Bin Cui, and Ming-Hsuan Yang.
\newblock Diffusion models: A comprehensive survey of methods and applications.
\newblock {\em arXiv preprint arXiv:2209.00796}, 2022.

\bibitem{Zdeborova2015}
Lenka Zdeborov{\'a} and Florent Krzakala.
\newblock Statistical physics of inference: Thresholds and algorithms.
\newblock {\em arXiv preprint arXiv:1511.02476}, 2015.

\end{thebibliography}

  \end{document}